\documentstyle[11pt,paspconf]{article}
\input{psfig.sty}

\begin{document}

\title{A 3D Treatment of Radiative Transfer \\ Including Multi-wavelength Scattering; Absorption and Far Infrared Emission; and Arbitrary Geometry}

\author{M.~Trewhella, B.~Madore and L.~Kuchinski}
\affil{IPAC, Caltech M.S. 100-22, Pasadena, CA 91125, USA}



\begin{abstract}

We have developed a new model that uses a cellular approach to calculate radiative transfer of starlight through dusty media. The model is designed to be user friendly enough to be distributed as a tool for use by the general astronomical community. Its features include treatment of scattering, absorption and far infrared emission; propagating many wavelengths simultaneously; giving 26 views of the output from different direction; and the ability to cope with arbitrary geometry.

As an example of the model's use, we have simulated the propagation of starlight through the edge-on galaxy NGC~891. As well as recovering the familiar edge-on view, we have a prediction for what the galaxy may look like if seen at other viewing angles.

\end{abstract}




\section{Introduction}
\subsection{Dust in disk galaxies}

Dust within the interstellar media of external galaxies undoubtably influences our view of the starlight. Exactly how much dust there is, how it is distributed, how much light it absorbs and how it affects our understanding of galaxy physics have been questions facing astronomers since the 1950s. 

\subsubsection{Galaxy brightness}

If a galaxy's starlight suffers greatly from its own internal extinction, then we will no measure its true brightness. Furthermore, the extinction correction is dependent on the viewing angle. The debate over extinction corrections (both absolute and inclination dependent) has become highly controversial - they were even omitted entirely from the RC3 catalogue (de Vaucoulers et al 1991).

\subsubsection{Galaxy morphology}

If the distribution of dust within a spiral is not totally uniform (as is likely) then extinction will not only reduce the galaxy's total brightness but change its appearance. The meaning of an apparent reduced surface-brightness at a given position is therefore ambiguous. It can either be due to less stars or more dust (Trewhella 1998).

\subsubsection{Star formation history}

Star formation history in galaxies is obtained by analysing the colour of the starlight. Dust extinguishes light of different wavelengths to a different degree, generally stronger towards the blue. Reddening by extinction can effectively disguise or even mimic a change in stellar populations, these two effects are not separable (de Jong 1996).

\subsubsection{Tully-Fischer relations}

It is quite possible that there are large differences of dust content and total extinction between galaxies. This will introduce a scatter in the Tully-Fischer relation. A possible way to minimise this is to use longer wavelengths that are less affected by dust. Giovanelli et al 1994, have shown a significant decrease in scatter can be achieved by using I band observations.  There is likely to be even more improvements made by moving to the near infrared wavelengths (e.g. K band) and by determining better extinction corrections.

\subsubsection{High redshift galaxies}

As we move to higher redshift, we are progressively observing shorter rest frame wavelength emission. This radiation is affected much more strongly by extinction. It is clear that if we are to gain useful information about the radiation density, star formation history and the epoch of galaxy formation, we are required to correct for dust extinction. It is only by scrutinising dust extinction in nearby galaxies that we can hope to extrapolate to distant, barely resolved sources.

\subsection{Radiative transfer models}

\subsubsection{Importance to extinction studies}

One of the major hindrances too our progress has been confusion over how to model the radiative transfer of the emitted starlight through the obscuring dust. Different models can produce dramatically different relationships between the dust optical depth (effectively dust mass) and the extinction.

The major factor contributing to these difference is the geometrical relationship between dust and stars. As a simple illustration, consider two hypothetical galaxies with the same amount of dust. In the first, the dust is placed in the mid-plane of the galaxy and in the second, the dust is placed as a screen in front of the galaxy. It is clear that the galaxy with the screen of dust will suffer more extinction per unit mass of dust. 

The other factor that has proved difficult to model is scattering. Scattering does not remove a photon's energy from a beam, it merely changes its direction of propagation. In the special case where dust only scatters and the dust and stars are distributed spherically symmetrically then scattering has no net effect on the observed surface-brightness of the object. As much light is scattered out of any given beam than is scattered into it. If the dust or star distributions are asymmetric, then starlight will be preferentially scattered into certain directions. This effect can be quite large, it is possible, for example, to construct a model where a disk seen face-on actually has a higher surface-brightness than if no dust where present.

Also, if a photon is scattered, then it will, on average, travel a longer path before escaping the galaxy. This increases the probability that it will be absorbed. Thus, neglecting scattering can lead to serious errors in predicting both the optical emission and the far infrared emission by dust heated via absorption.

\subsubsection{History}

The first radiative transfer models came about in the 1930s and dealt with extinction on the line of sight to stars in our Galaxy. This specialised situation is relatively simple to model - both scattering and absorption only remove energy from the beam and all the light has to pass through all the dust. 

External galaxies were first addressed by Holmberg (1959). He effectively compared the surface-brightness of a sample of galaxies seen at different inclinations. Interpreting this data with a screen model, gives the conclusion that galaxies are optically thin. This result was confirmed by later surface-brightness/inclination tests and became the accepted view of galaxies.

Two radiative transfer models of galaxies that were many times more sophisticated were developed in the late 1980s (Kylafis and Bachall 1987, Bruzual et al 1988). These models both included the effects of scattering and considered more realistic geometries. It was not, however, until Disney et al. (1989) investigated analytical models (without scattering) of several types of realistic galaxy geometries in the context of surface-brightness/inclination tests that the full significance of the choice of radiative transfer model became apparent. They found that Holmberg's original data could be interpreted to show almost any optical depth by changing the geometrical parameters. Indeed, with the most realistic models, it seemed that galaxies might well be optically thick!

In the early 1990s, extinction in galaxies became a highly controversial subject, culminating in a NATO Advanced Research Workshop in the subject in Cardiff in 1994 (Davies and Burstein 1995). During this meeting, it was universally agreed that the extinction problem would not be solved without the use of full-scale realistic 3D radiative transfer models. The criteria suggested for an adequate model were absorption; scattering; clumpiness; and far infrared emission by dust.

\subsubsection{Modern models}

The Monte-Carlo approach where individual photons are randomly created and followed though their paths is by far the most popular approach (Witt et al. 1992, Bianchi et al. 1996, van Buren et al. 1998). Their main advantage is that scattering can be treated simply and logically, the imaginary photon's path mimicking that taken by real photons. They can also, in principle, handle arbitrary geometry. In practice, they have only recently been used to investigate extinction in clumpy geometries (Witt and Gordon 1996). So far, these models have only been used to investigate general effects of dust on starlight in galaxies and not yet been applied to real galaxies.

The alternative model uses an analytical approximation in the radiative transfer equation to allow all the energy scattered into a given direction at a given point to be calculated from the first two terms (Kylafis and Bahcall 1987). This model does not consider clumpiness but has been fitted to images of two edge-on galaxies (where the clumpiness is smoothed out), UGC 2048 (Xilouris et al. 1997) and NGC 891 (Xilouris et al. 1998) with impressive results.

\section{Cellular model}

The models described above represent an enormous advance and go a long way to realising the goals made apparent by the Cardiff dust meeting. They each have strengths and weaknesses, however, and it is likely that one model may be more suitable than the other to simulate a given situation. With this in mind, it is prudent to investigate whether there exist other models that may be more suitable for certain tasks. In designing our model we added two more criteria; that it be flexible enough for a wide range of applications - not necessarily just galaxies; and that it be user friendly enough to allow non-specialists to take advantage of it.

\subsection{Model design}

The theoretical concept is to divide a given 3D space up into cubic cells or crystals. Radiation is envisaged as residing on the surface of the 26 'facets' that represent the interface between the crystal and its touching neighbours (the 6 cardinal directions, 12 edges and 8 corners). Radiation is passed from a facet into the adjacent crystal where it interacts with the dust contained in the crystal. The amount of optical energy absorbed by the dust is calculated from the optical depth and the albedo. This absorbed energy is then redistributed equally among the 26 facets of the new crystal as far infrared radiation. The amount of optical energy allowed to pass straight through the crystal is calculated from the optical depth. The amount of optical energy scattered into the remaining 25 directions is calculated from the optical depth, the albedo, the scattering angle and the scattering asymmetry factor. This is done for each wavelength before moving on to the next crystal.

The model transfers energy from the brightest crystal first and then moves on to the next brightest. Gradually, the energy accumulates on the exterior facets of the outermost crystals. The simulation ends when a user specified fraction of the total energy is contained on these exterior facets. This energy is then read out to produce one image per wavelength (plus a far infrared) from each of the 26 directions specified by the exterior facets.

\subsection{Model use}

\subsubsection{Input}

The model takes two plain text input files, one describing the dust and star distribution and one describing the dust properties.

The format for the dust properties input file is;
\begin{verbatim}
l   wave    w    g
\end{verbatim}
where l is an integer wavelength identifier, wave is the wavelength, w is the albedo and g is the scattering asymmetry factor.

The format for the dust star input file is;
\begin{verbatim}
x   y   z   l   E   tau
\end{verbatim}
where (x,y,z) are integers specifying the 3D position, l is the integer wavelength identifier corresponding to that given in the dust properties input file, E is the energy contained within the crystal and tau is the optical depth.

The input files can be created in many different ways, ranging from a automated program to directly typing in by hand. This allows complete freedom for the user to specify arbitrary geometry.

\subsubsection{Output}

The output of the model is in the form of data files similar to CCD images. The files are named {\tt wnnfmm.dat}. Where nn is the integer wavelength identifier and mm is the facet number (equivalent to viewing angle). There is be 26 files for each input wavelength and a further 26 for the far infrared images. The format for the data files is;
\begin{verbatim}
i    j     I
\end{verbatim}
where (i,j) are integers specifying the position on the image and I is the intensity. These data files can be either by analysed directly (using IDL, SuperMongo, PGplot etc) or read into a standard image analysis package. The major packages, IRAF, STARLINK and MIDAS can all convert data from this format. The images can then be analysed in the same way as regular CCD images.

\subsection{Potential applications}

In general, the model can be used for many applications where emitted energy undergoes some arbitrary interaction before escaping. The interactions need not even be with dust, as long as the optical properties of the intervening material can be specified. For example, to simulate scattering by electrons the albedo would be set to 1 - purely scattering.

The part of the code that contains the physics of the radiative transfer is only 10 lines long. This could easily be replaced with a physical description of another kind of radiation-particle interaction.

\section{Application to disk galaxies}

The application that inspired the building of this model is radiative transfer in disk galaxies. Disk galaxies have extremely complex geometries and if we are to fully understand the radiative transfer, we need to be able to include a smooth component as well as GMCs, spiral arms and star formation regions. The cellular model can deal with arbitrary distributions as easily as ones with analytical descriptions, it is therefore relatively simple to simulate complete spirals.

\subsection{NGC 891?}

Xilouris et al. 1998, have fitted their 3D radiative transfer model to the edge-on galaxy NGC~891. Table~2 of their paper gives the derived parameters for their fit. We will use this description of NGC~891 as the input to our radiative transfer model. In addition, we will add random clumpiness and spiral arms to the model. This will demonstrate the ability of our model to simulate complex geometry and give images from many viewing angles.

\begin{figure}
\centerline{\psfig{file=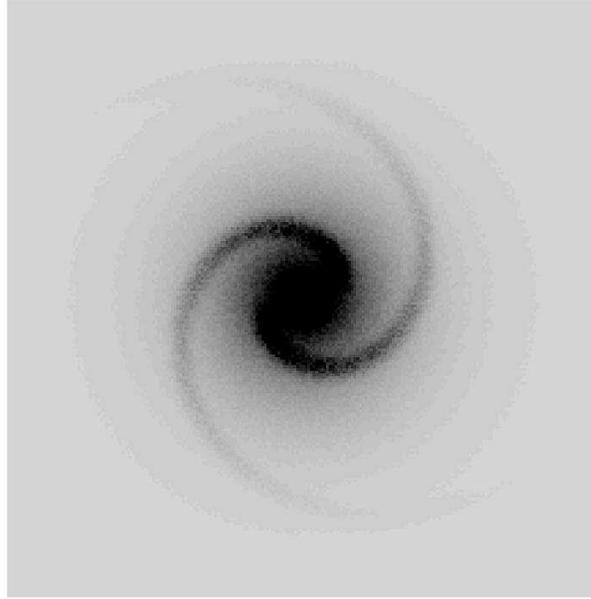,height=80mm}}
\caption{Face-on view of NGC~891? This is the central plane ($z=0$) of a 3-D energy distribution chosen to represent NGC~891 in the V band. The exponential disk and bulge were taken from Xilouris et al. 1998 and the spiral arms were added for effect.} \label{fig:2d_opt}
\end{figure}

Figure~\ref{fig:2d_opt} shows a slice through the central plane of the input intensity. The spiral arms were chosen to look reasonably realistic in this view. How representative is this of how NGC~891 would appear if viewed face-on? It is not possible to answer this question without summing all the energy from the cells other than in the central plane and taking account of the dust.

\begin{figure}
\centerline{\psfig{file=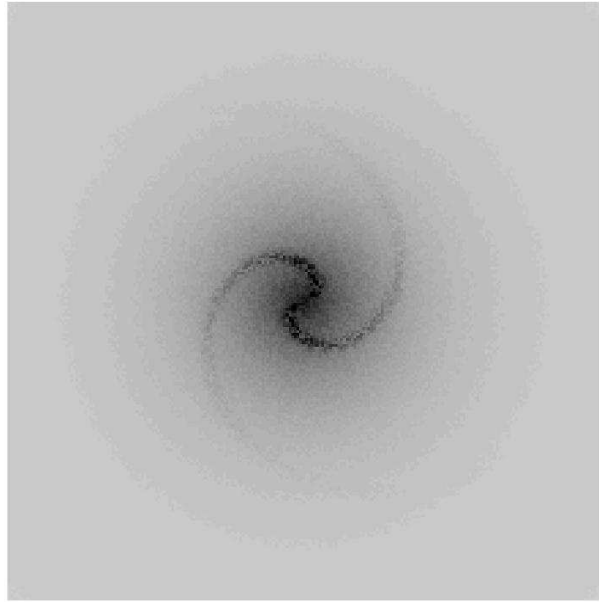,height=80mm}}
\caption{Dust distribution of NGC~891? This is the central plane ($z=0$) of a 3-D dust distribution chosen to represent NGC~891 in the V band. The exponential scale lengths were taken from Xilouris et al. 1998 and the spiral arms were added for effect. The spiral arms are slightly narrower and offset from the optical arms.} \label{fig:2d_tau}
\end{figure}
 
Figure~\ref{fig:2d_tau} shows a slice through the central plane of the dust input optical depth. The spiral arms were chosen to be narrower and slightly offset (inwards) from the optical arms.

A $(110\times 110\times 9)$ array of intensity and optical depth was defined using the 3D V band data described above. This was used as the input to the radiative transfer model along with standard V band dust properties of $\lambda = 0.44$, $\omega = 0.66$ and $g = 0.59$.

\begin{figure}
\centerline{\psfig{file=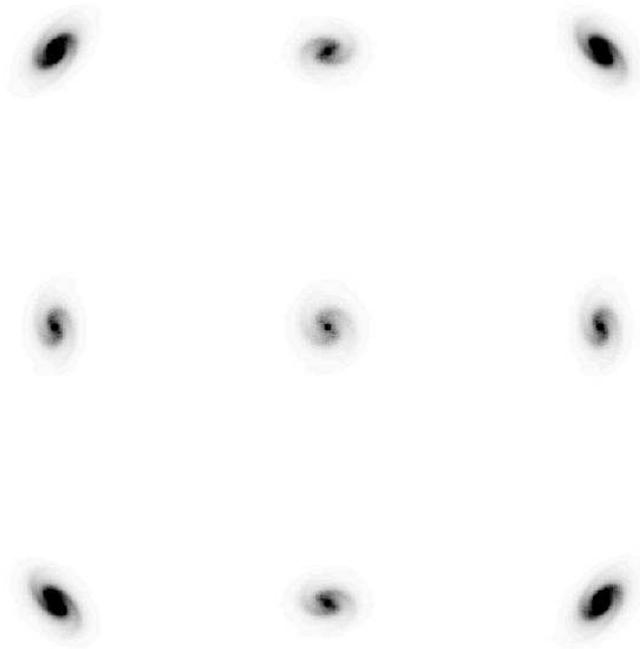,height=110mm}}
\caption{V band model output of suggested NGC~891 simulation. The central image is the face-on view, the other images are of the views adjacent to face-on.} \label{fig:w01f03}
\end{figure}

\begin{figure}
\centerline{\psfig{file=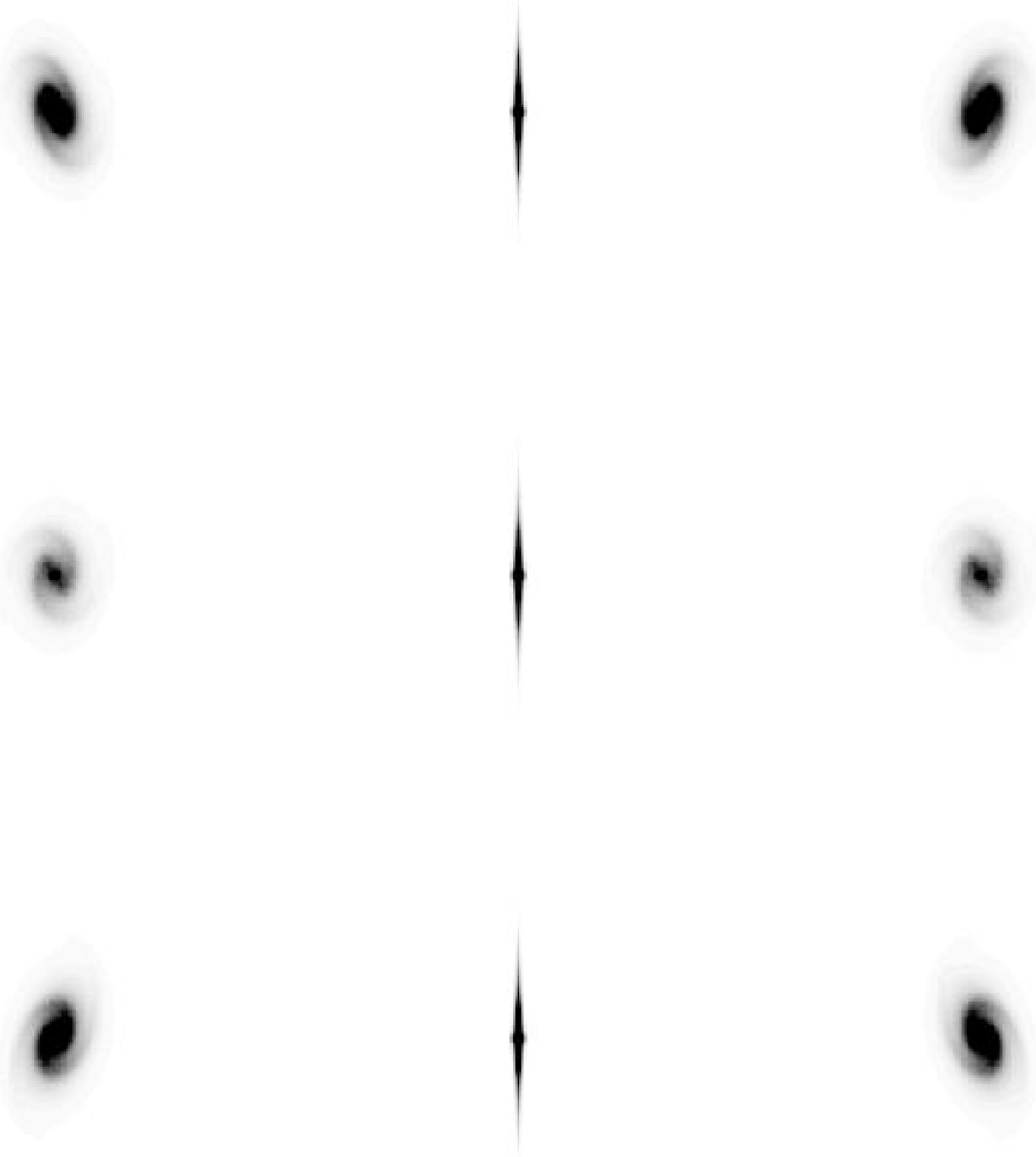,height=110mm}}
\caption{V band model output of suggested NGC~891 simulation. The central image is the edge-on view, the other images are of the views adjacent to edge-on.} \label{fig:w01f02}
\end{figure}

\begin{figure}
\centerline{\psfig{file=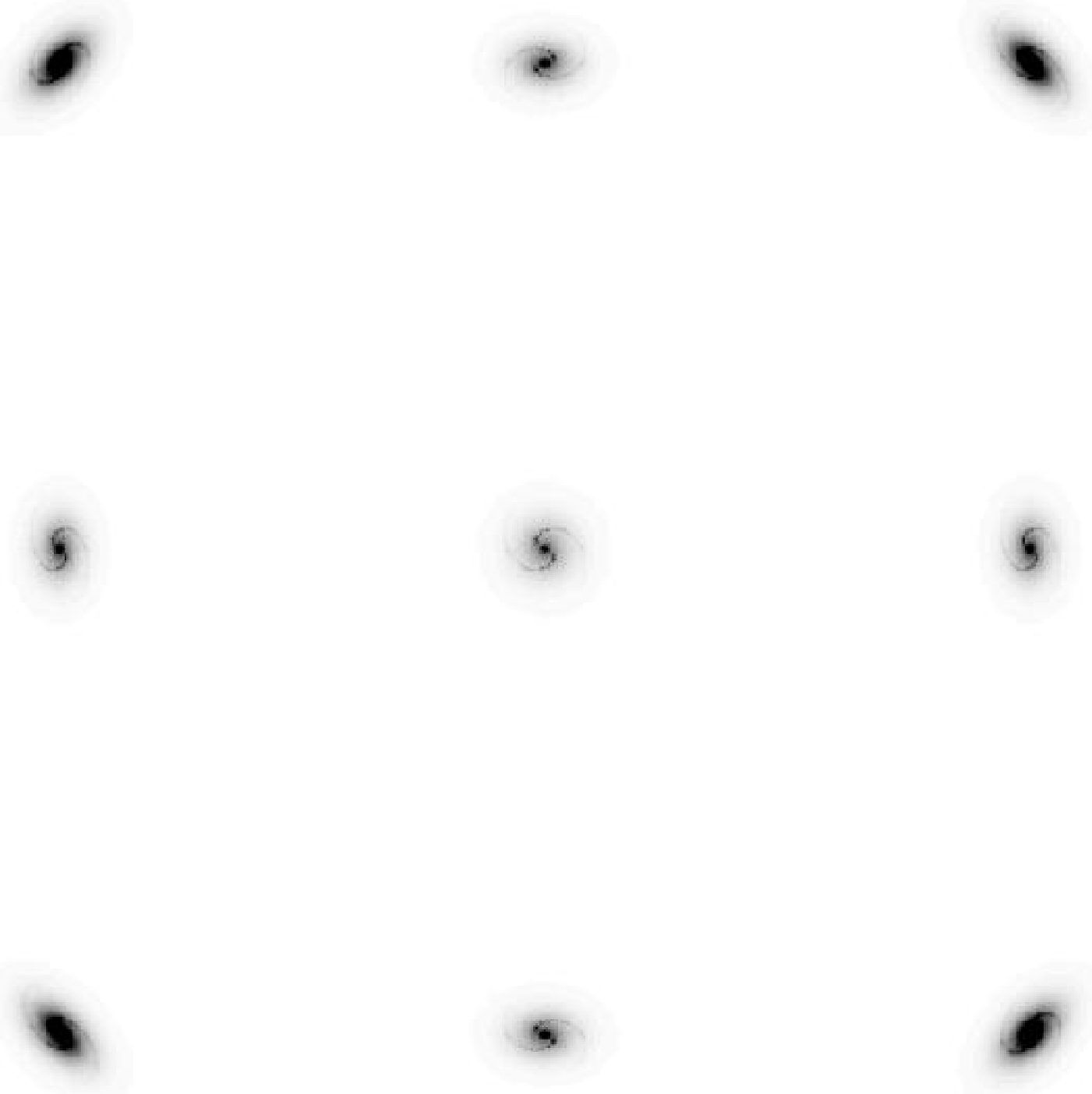,height=110mm}}
\caption{Far infrared model output of suggested NGC~891 simulation. The central image is the face-on view, the other images are of the views adjacent to face-on.} \label{fig:w00f03}
\end{figure}

\begin{figure}
\centerline{\psfig{file=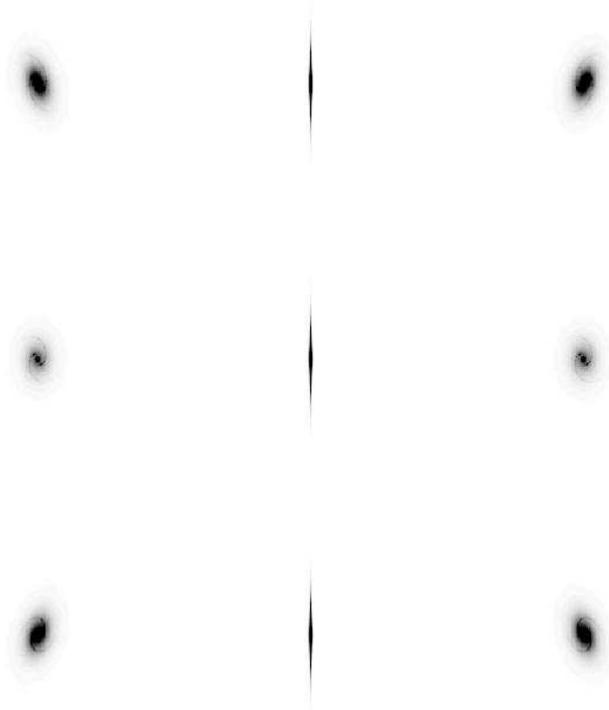,height=110mm}}
\caption{Far infrared model output of suggested NGC~891 simulation. The central image is the edge-on view, the other images are of the views adjacent to edge-on.} \label{fig:w00f02}
\end{figure}

The results are shown in figures~\ref{fig:w01f03},~\ref{fig:w01f02},~\ref{fig:w00f03}~and~\ref{fig:w00f02}. Each of the figures shows 9 views of the output. Figure~\ref{fig:w01f03} has the face-on view in the centre and the adjacent 4 edges and 4 corners around the edge. To view the outside images with the correct orientation relative to the central, rotate the figure until the image you wish to view is at the bottom. The image is the representative of what you would see by rotating vertically from the central image. Figure~\ref{fig:w01f03} thus shows 9 different views of NGC~891 from the more familiar edge-on - impossible to do in real life! Figure~\ref{fig:w01f02} has the edge-on view in the centre with its adjacent viewing angles around the outside. The equivalent face-on and edge-on views of the dust emission are given in figures~\ref{fig:w00f03}~and~\ref{fig:w00f02} respectively. Of course, no far infrared images as yet have the resolution to test these images but they should be available soon with the advent of the sub-millimetre bolometer arrays SHARC and SCUBA on the CSO and JCMT respectively.

\section{Future}

\subsection{Galaxies}

We have imaging photometry of 10 nearby late-type galaxies from the vacuum ultra-violet through to the sub-millimetre. This data a hugh amount of information about the dust and star distributions. When we have tailored our models to match all the data, we can be pretty confident that we have a realistic description of the galaxy's structure. We will be able to use this information to gain information about the star formation and its history at all positions in our 3D description.

We will search the sample of 10 galaxies for trends differences and more simple indicators of extinction. The many views of each galaxy will allow us to determine empirical extinction corrections that span a range of galaxy types and physical conditions.

The effects of high redshift on the models can be simulated, giving us a more accurate way to determine intrinsic spectra, brightnesses and star formation rates of high redshift galaxies.

\subsection{Release}

We intend to release the model to the astronomical community as a general tool for radiative transfer as soon as we finish testing and verifying its operation. To be added to the mailing list for information of how and when this will be done email;
\begin{verbatim}
mxt@ipac.caltech.edu
\end{verbatim}

\subsection{Improvements}

We are currently considering 3 improvements.

\begin{enumerate}

\item {\bf Dust temperature.} At the moment, all the far infrared emission by the heated dust is considered as a single unit. We intend to upgrade the code to calculate dust temperatures and therefore the far infrared emission at different wavelengths.

\item {\bf Far infrared opacity.} At the moment, the far infrared emission is treated as being essentially optically thin. This may not be the case for some astrophysical situations. When the far infrared emission is split into separate wavelengths, far infrared opacity will also be added.

\item {\bf Dust property variation.} At the moment, the dust properties are fixed throughout the whole 3D space. It may be desirable to have different dust properties in different parts of the simulation.

\end{enumerate}

Other improvements will be considered if potential users require them.

\section{Summary}

We have developed a new model that uses a cellular approach to calculate radiative transfer of starlight through dusty media. The model is designed to be user friendly enough to be distributed as a tool for use by the general astronomical community. Its features include treatment of scattering, absorption and far infrared emission; propagating many wavelengths simultaneously; giving 26 views of the output from different direction; and the ability to cope with arbitrary geometry.

\acknowledgments

We would like to acknowledge the extreme usefulness of discussions with other radiative transfer model builders and dust specialists including Adolf Witt, Simone~Bianchi, Manolis~Xilouris, Nick~Kylafis, Jon~Davies, Paul~Alton, Dave~van-Buren and Michael~Brundage.


%
%

%

\end{document}